# An Algebraic Approach for Computing Equilibria of a Subclass of Finite Normal Form Games

Samaresh Chatterji and Ratnik Gandhi

Dhirubhai Ambani Institute of Information and Communication Technology,
Post Bag No. 4, Gandhinagar 382007, India
`{samaresh_chatterji,ratnik_gandhi}@daiict.ac.in`

**Abstract.** A Nash equilibrium has become important solution concept for analyzing the decision making in Game theory. In this paper, we consider the problem of computing Nash equilibria of a subclass of generic finite normal form games. We define *rational payoff irrational equilibria games* to be the games with all rational payoffs and all irrational equilibria. We present a purely algebraic method for computing all Nash equilibria of these games that uses knowledge of Galois groups. Some results, showing properties of the class of games, and an example to show working of the method concludes the paper.

## 1 Introduction

The problem of computing all equilibria of class of integer payoff irrational equilibria(IPIE) games is studied in [1]. The approach in [1] is as follows: They characterize equilibria of an IPIE game as solutions to a system of polynomial equations called a *game system* $\mathcal{GS}$. With a numerical method they compute a sample solution of the $\mathcal{GS}$ and follow it up by the Galois group action for computing all equilibria solutions. Existing methods for computing equilibrium, such as the approach based on the homotopy continuation methods, given in [2], have the drawback of providing solutions via approximation. On the other hand the method in [3] is highly dependent on a chosen probability distribution.

With the motivation and game model presented in [1], in this work, we extend the approach of using Galois group for computing all equilibria of the IPIE games to a larger class of rational payoff irrational equilibria(RPIE) games. The approach is a purely algebraic and provides a marginal decrease in computational time compared to the method based only on Gröbner basis [4]. The method computes exact equilibria and does not depend on probability distributions. We further show that for a subclass of games it computes equilibria in closed form.[1]

The number of equilibria in a finite normal form game increases exponentially in the size of the game [2]. This means any method for computing all equilibria, one at a time, are bound to be exponential. In the light of this result structural relations between equilibria solutions of a game can help us compute all equilibria in a better way. In this paper, our aim is to present a pure algebraic method that computes all equilibria of RPIE games using knowledge of its sample equilibrium. Thus, answering partially an important question of establishing relations between all equilibria solutions, raised by McKelvey and McLennan [5].

To our knowledge, a method for computing equilibria of RPIE games with Galois groups has not been considered earlier.

---

[1] cf. Proposition 3.

Organization of the article is as follows. Preliminaries and the game model is presented in Section 2. The algorithm for computing all equilibria of RPIE games and related results are presented in Section 3. Section 4 contains a worked out example of RPIE game.

## 2  Preliminaries

The subclass of games that we consider are generic and are known to have finitely many odd number of irrational equilibria [6]. Our convention is to write totally mixed real-irrational Nash equilibria as irrational Nash equilibria.

**Definition 1.** *A finite normal form game with*

- *its all payoff values rational numbers*
- *its all Nash equilibria irrational numbers*

*is called an rational payoff irrational equilibria(RPIE) game.*

The characterization of equilibria as solutions to a system of polynomial equation and the underlying game model, that we consider in this paper, is based on [1].

Let $T$ be an RPIE game with $n = |N|$ players. Each player $i$ has $k_i \geq 2$ strategies, $|S_i| = k_i$. We write $\mathcal{K}^+ = \sum_{i=1}^n k_i$. $A^i_{j_1 j_2 \ldots j_n}$ denotes the payoff received by player $i$ when each player adopts strategy $j_m$ for $1 \leq j_m \leq k_m$ and $m = 1, \ldots, n$. The probability that player $i$ chooses strategy $j_i \in \{1, 2, \ldots, k_i\}$ is denoted by $x^i_{j_i}$. By Definition 1 an RPIE game has

$$0 < x^i_{j_i} < 1. \tag{1}$$

Moreover, for each player $i$,

$$\sum_{j_i=1}^{k_i} x^i_{j_i} = 1. \tag{2}$$

Expected payoff for player $i$,

$$\alpha_i = \sum_{j_1=1}^{k_1} \sum_{j_2=1}^{k_2} \ldots \sum_{j_n=1}^{k_n} A^i_{j_1 j_2 \ldots j_n} x^1_{j_1} x^2_{j_2} \ldots x^n_{j_n} \tag{3}$$

In a Nash equilibrium, the following holds:

$$\alpha_i \geq \sum_{j_1=1}^{k_1} \sum_{j_2=1}^{k_2} \ldots \sum_{j_{i-1}=1}^{k_{i-1}} \sum_{j_{i+1}=1}^{k_{i+1}} \ldots \sum_{j_n=1}^{k_n} A^i_{j_1 j_2 \ldots j_{i-1} j_i j_{i+1} \ldots j_n} x^1_{j_1} x^2_{j_2} \ldots x^{i-1}_{j_{i-1}} x^{i+1}_{j_{i+1}} \ldots x^n_{j_n},$$
$$\text{for every } j_i \in S_i \text{ and for every } i \in \{1, \ldots, n\}. \tag{4}$$

With (4), (2) and (1) we get a system of polynomial equation of the form.

$$\alpha_i - \sum_{j_1=1}^{k_1} \sum_{j_2=1}^{k_2} \ldots \sum_{j_{i-1}=1}^{k_{i-1}} \sum_{j_{i+1}=1}^{k_{i+1}} \ldots \sum_{j_n=1}^{k_n} A^i_{j_1 j_2 \ldots j_{i-1} j_i j_{i+1} \ldots j_n} x^1_{j_1} x^2_{j_2} \ldots x^{i-1}_{j_{i-1}} x^{i+1}_{j_{i+1}} \ldots x^n_{j_n} = 0$$
$$\text{for every } j_i \in S_i \text{ and } i \in \{1, \ldots, n\}. \tag{5}$$

For the purpose of computing all equilibria an RPIE game we shall consider the system of polynomial equations of the form (5) and call it a *game system* $\mathcal{GS}$. We shall be specifically interested in the following field arithmetic for the $\mathcal{GS}$ representing RPIE games.

**Definition 2.** *Let $K \supset F$ be a finite extension of a field $F$. Then the Galois group $G = Gal(K/F)$ is the set*

$$G = \{\sigma : K \to K | \sigma \text{ is an automorphism, } \sigma(a) = a \text{ for all } a \in F\}.$$

We now consider the particular situation when $K$ is an extension of $F$ of the form $F(\alpha)$ where $\alpha$ is a root of a polynomial $p(x) \in F[x]$. It is known that the Galois group $G$ of the extension $K$ acts as a permutation group on the roots of the polynomial $p(x)$.

In this work our assumption is that these Galois groups are known, and we further utilize them for computing solutions of the polynomial system. It is important to note that for computing a Galois group of $n!$ elements, computation of its $2n$ generators is sufficient [7]. It is a procedure not requiring exponential computation . And so the assumption, that the Galois groups are known, is feasible. The polynomials that we consider are multilinear – in indeterminate variables – over the field of rational numbers $\mathbb{Q}$ and generate irrational equilibria solutions. Thus the polynomial system induces non-trivial Galois groups.

## 3 Equilibria of RPIE Games

In this section, we outline the algorithm to compute all irrational equilibria of an RPIE game. We further show correctness of the algorithm and present other results.

### 3.1 Method

Before formally presenting our algorithm, we briefly discuss the approach and the underlying assumptions. We assume that we have an RPIE game $T$. As described in Section 2, we can derive a system of polynomial equations $\mathcal{GS}$ whose solutions include all the Nash equilibria of the game $T$. Since some of the solutions need not be Nash equilibria, our algorithm rejects these unwanted solutions using various different mechanisms. Due to Bernstein's theorem [8], we get an upper bound on the number of solutions a polynomial system can have. Bounds on the number of equilibria solutions – out of all polynomial solutions – are available in [9,10]. An upper bound on the number of solutions and bounds on the number of equilibria solutions give a bound on the number of solutions to be computed and the number of non-equilibria solutions to be rejected.

In the initial phase of our method, Buchberger's algorithm is called to derive a univariate polynomial in the Gröbner basis(GB) of the $\mathcal{GS}$.[2] Since the game is RPIE, Nash's theorem [12] guarantees that the univariate polynomial has at least one irrational root. A root of the univariate polynomial is computed and substituted in the triangular form of a Gröbner basis to find a univariate polynomial in other indeterminate variables. We repeat this procedure at most $\mathcal{K}^+ - n$ times, and at the end of it we have an irrational solution of the $\mathcal{GS}$, a sample solution.

We denote a Galois group of the irreducible part of a univariate polynomial in GB of $\mathcal{GS}$ by $G$. We assume $G$'s are known.

---

[2] Recall that the system of polynomial equations $\mathcal{GS}$ over complex number field has finitely many solutions. This finite variety of the $\mathcal{GS}$ (or equivalently zero-dimensional ideal $\mathcal{I}$ of the $\mathcal{GS}$) guarantees a univariate polynomial in its Gröbner basis [11].

In the next phase, we apply the transitive Galois group action corresponding to each indeterminate variable and find Galois-orbits to determine all irrational solutions of the $\mathcal{GS}$.

The final phase consist of determining all non-equilibria solutions and rejecting them. For this we use the Nash equilibrium verification algorithm in [13]. Following is an outline of the algorithm to compute all equilibria of an RPIE game with Galois groups.

---

**Algorithm 3.1** Computing All Nash Equilibria of an RPIE game.

Input: An RPIE game, Galois groups.
Output: All equilibria of the input RPIE game.

1: $\beta = (\beta_1, \beta_2, \ldots, \beta_{\mathcal{K}+})$.                    {Initialize an empty tuple to store a sample solution of the $\mathcal{GS}$}.
2: Characterize all the Nash Equilibria of the input game as solutions to the $\mathcal{GS}$.
3: Call Algorithm 3.2 with $\mathcal{GS}$ for computing a sample equilibrium of the input RPIE game.
4: Call the Galois group action Algorithm 3.3 with the sample solution tuple saved in $\beta$.
5: Save output of the Algorithm 3.3 in $X$.
6: Reject non-equilibria solutions of the $\mathcal{GS}$ from $X$ using verification algorithm in [13] or criteria (1) and (2).

---

**Algorithm 3.2** Computation of a sample solution.

Input: $\mathcal{GS}$ of the input game.
Output: A sample solution $\beta = (\beta_1, \beta_2, \ldots, \beta_{\mathcal{K}+})$ of the input game.

1: With Buchberger's Algorithm on $\mathcal{GS}$, compute triangular form of GB.
2: **while** one sample solution $\beta$ of the $\mathcal{GS}$ is not constructed **do**
3:     Compute a root $\alpha$ of univariate polynomial – of some indeterminate variable $x_i$ – generated in Step 3.
4:     **if** $\alpha \in \mathbb{Q}$ **then**
5:         Reject $\alpha$ and go to Step 3.
6:     **else**
7:         Save $\alpha$ in $\beta$ at location $\beta_i$.
8:     **end if**
9:     Substitute the root $\beta_i$ in $\beta$ into $\mathcal{GS}$ and compute a new triangular form with one less indeterminate variable.
10: **end while**

---

The group action algorithm is given in [1]. We reproduce it here for the discussion to be complete. Algorithm 3.3 computes group action by transitive Galois groups. The action is computed for each indeterminate variable $x_i$ by considering it over each coordinate root in the tuple $\beta$. The action generates Galois-conjugates of the roots that are further saved in as solution tuples in the set $X$.

Traditional approach, given in [4], for computing solutions of system of polynomial equations using GB calls the Buchberger's algorithm for computing a triangular form. The triangular form provides a univariate polynomial in one indeterminate variable. Each root of the univariate polynomial is then substituted back in the triangular form to compute corresponding solution tuple, requiring multiple substitutions and factorizations. Algorithm 3.2, on the other hand, invokes Buchberger's algorithm exactly once. The Algorithm

**Algorithm 3.3** Computing orbit of a Galois Group Action.
Input: A sample solution $\beta$ of the $\mathcal{GS}$.
Output: All the conjugate solutions of the input sample solution in set $X$.

1: Initialize the processed-elements list $X$ and unprocessed-elements list $U$ as $X = U = \{\beta\}$.
2: **while** $U$ is not empty **do**
3:   Let $u = (u_1, u_2, \ldots, u_{\mathcal{K}^+})$ be the first element of $U$. Delete $u$ from $U$.
4:   **for** each $i$ and $j$, $g_j^i$ in Galois group $G_i$ and $u_i \in u$. **do**
5:     Compute the transitive Galois group action $u_i^{g_j^i}$.
6:     $\beta' = (u_1^{g_j^1}, u_2^{g_j^2}, \ldots, u_{\mathcal{K}^+}^{g_j^{\mathcal{K}^+}})$.
7:     **if** $\beta' \notin X$ **then**
8:       $X = X \cup \{\beta'\}$ and $U = U \cup \{\beta'\}$.
9:     **end if**
10:   **end for**
11: **end while**

3.2 computes a sample solution tuple corresponding to first irrational root of the univariate polynomial. Rest of the solutions are then generated by polynomial time group action, requiring no further substitutions and factorizations.

It is important to note that, due to [12] and the fact that the input game has all irrational equilibria, we are guaranteed to get one solution of the $\mathcal{GS}$ in $\beta$ and so the Algorithm 3.1 reaches Step 4, every time. Next it calls the Algorithm 3.3 for computing polynomial time Galois group action over available sample solution in the $\beta$. In the Algorithm 3.3 all other conjugate roots are computed with their known Galois groups $G_i$.

Moreover, finite group action on finite variety guarantees that the Algorithm 3.3 reaches Step 13. At the end of Step 13, Algorithm 3.3 generates solutions of polynomial system $\mathcal{GS}$ in $X$, all of which may not be Nash equilibria. We use polynomial time algorithm, suggested in [13], to reject the non-equilibria solutions.

For checking the condition in Step 4 of the Algorithm 3.2, i.e., for deciding $\alpha \in \mathbb{Q}$ we consider the following approach. Compute an approximation of $\alpha$ with suitable numerical method. With the KLL Algorithm [14] compute the minimal polynomial for $\alpha$. Check irreducibility of the minimal polynomial over $\mathbb{Q}$ with algorithm in [15]. The irreducibility test decides whether $\alpha \in \mathbb{Q}$ or not.

The Algorithm 3.2 computes a sample solution of the $\mathcal{GS}$. Various approaches for computing a sample equilibrium of a game are discussed in [5]. The approach that we consider of using Gröbner basis, though not new, is developed independently. Contrary to the other approaches based on Gröbner basis, our approach, given in Algorithm 3.2, differs primarily in Step 4.

The Algorithm 3.1 shows a method for computing solutions of a system of polynomial equations without having to factorize the system every time.

### 3.2 Comparison with the Algorithm in [1]

The Algorithm in [1] calls MVNRM for computing an approximate solution of the $\mathcal{GS}$. MVNRM is followed by the KLL algorithm for converting an approximate root in to algebraic form. The solution is then tested for irrationality. In case otherwise, a new approximate solution is computed.

On the other hand, the Algorithm 3.1 uses Buchberger's algorithm for computing a Gröbner basis(GB). This converts the $\mathcal{GS}$ in triangular form, requiring doubly exponential time in $\mathcal{K}^+$. A univariate polynomial in the triangular form is factorized to find an irrational root. The root is substituted back in the GB for finding a complete irrational solution. This requires rejection of rational roots, if any.

The difference in both the methods is: the Gröbner basis method is purely algebraic and allows lot more structure to manipulate $\mathcal{GS}$. While the numerical MVNRM provides lot more efficiency, disclosing a little about the structure of the polynomials, and interrelations of its solutions.

### 3.3 Results

The Algorithm 3.1 computes all equilibria of RPIE games with $n \geq 3$ players. Following results show, why it can not work for games with $n = 2$ players.

**Proposition 1.** *A bimatrix game with all rational payoff values has all rational equilibria.*

*Proof.* For computing all equilibria of a bimatrix game, system of linear equations are sufficient [16]. That is, if all the game payoff values are defined over field $F$, then all of its equilibria – characterized as solutions to system of polynomial equations – can be found in the field $F$ only.

□

Following is an immediate corollary to the result above.

**Corollary 1.** *Algorithm 3.1 can not be used to compute equilibria of a bimatrix game defined over arbitrary field.*

*Proof.* From Proposition 1, it is clear that the class of two player games do not produce field extensions. This means that corresponding Galois groups are trivial, i.e., $Gal(\mathbb{F}/\mathbb{F}) = \{e\}$. This proves the claim. □

**Corollary 2.** *The class of RPIE games is empty for $n = 2$ players.*

*Proof.* Follows from Definitions 1, 2 and Proposition 1. □

Next result shows correctness of the Algorithm 3.1.

**Proposition 2.** *The Algorithm 3.1 to compute equilibria of the class of RPIE games works, i.e., the Algorithm 3.1 generates all irrational equilibria and no other solutions at termination.*

*Proof.* An input RPIE game $T$ with $n \geq 3$ players is characterized as the $\mathcal{GS}$ of form (5). Polynomial system comes from the inequality on expected payoffs and payoffs at pure strategies. This causes the $\mathcal{GS}$ to have more solutions then just the equilibria.

In the first phase, Algorithm 3.1 calls the Algorithm 3.2. The algorithm 3.2 computes a sample solution $\beta$ by first building a Gröbner basis for the $\mathcal{GS}$ using Buchberger's algorithm. Buchberger's algorithm terminates with triangular form analogous to echelon form in the linear case.

Since the game is known to be RPIE and rational solutions of the $\mathcal{GS}$ are rejected by the Algorithm 3.2, the sample solution $\beta$ must have all irrational coordinates. Consequently, each coordinate $\beta_i$ of the sample solution $\beta$ results in an algebraic extenion $K = \mathbb{Q}(\beta_i)$ of $\mathbb{Q}$ with finite Galois group $G_i = Gal(\mathbb{K}/\mathbb{Q})$. Since the group action of $G_i$ is transitive, it generates all irrational solutions of the $\mathcal{GS}$.

Zero-dimensional ideal of $\mathcal{GS}$ guarantees that the group action terminates. This enables the Algorithm 3.1 to reach Step 5, every time there is a RPIE game $T$ as its input. The algorithm generates solutions of the $\mathcal{GS}$ that contain all the equilibria of the game $T$.

Finally, Algorithm 3.1 rejects solutions of the $\mathcal{GS}$ which are not Nash equilibria. Since the set of Nash equilibria is known to be non-empty, set $X$ contains all and only the Nash equilibria solutions of the RPIE game $T$. □

With the available finite precision technology for representing a number in computer memory, the problem of storing an irrational equilibria is important [17]. We show that the issue for RPIE games can be resolved as follows.

**Proposition 3.** *If univariate polynomials in ideal $\mathcal{I}$ of $\mathcal{GS}$ of an RPIE game has solvable Galois group. Then the Algorithm 3.1 computes Nash equilibria of the game in closed form.*

*Proof.* Due to Galois correspondence, we know, if a polynomial has solvable Galois group, all its roots can be computed by radicals. If each univariate polynomial in ideal $\mathcal{I}$ of $\mathcal{GS}$ of an RPIE game has solvable Galois group,[3] then roots of this set of polynomials can be computed using radicals. This gives all solutions in closed form, a subset is Nash equilibria of the game. □

It is known that all abelian groups, groups of order $< 60$, group with odd order (Feit-Thompson theorem) and groups of order $p^a q^b$ are solvable, where $p$ and $q$ are prime. Moreover, some non-abelian group may also be solvable [19]. This suggests that Proposition 3 is applicable for a substantial number of games. The equilibria of games with non-solvable Galois groups can be obtained in algebraic form by first computing equilibria solutions numerically, and then constructing minimal polynomial for each of the numerical values with the algorithm in [14].

## 4  Example

In this section, we show working of Algorithm 3.1 with an example of 3 players 2 strategy RPIE game. With the Membership algorithm in [1], we verify that the game, given in Table 1 is RPIE.

|   | A | B |   | A | B |
|---|---|---|---|---|---|
| a | 6, -1, 4 | 0, 9, 0 | a | 2, 0, 0 | 0, 9/2, 0 |
| b | 0, 3/2, 0 | 2, 0, 0 | b | 0, 27/2, 0 | 4, 0, 6 |
|   | 1 |   |   | 2 |   |

**Table 1.** Payoff table of a 3 player 2 strategy RPIE game. Player 1 and 2's strategies are indicated by a, b and A, B respectively. Player 3's strategies are 1 and 2. Entry in each cell of payoff table indicates player 1, 2 and 3's respective payoff for their respective strategies.

We let $x = x_1^1, y = x_2^1, z = x_3^1$ be the first strategy of players 1, 2 and 3 respectively. Probability that players will choose their second strategy is $1-x, 1-y$ and $1-z$ respectively. After characterizing Nash equilibria as the solutions of a $\mathcal{GS}$ of form (5), we apply Buchberger's algorithm, with $x \prec y \prec z$, to compute a univariate polynomial $5x^2 - 27x - 27 = 0$ in a GB. The univariate polynomial has $x = \frac{3}{10}(-9 \pm \sqrt{141})$ as its two roots, and is irreducible over $\mathbb{Q}$. Its Galois group ({id,conjugate}) is isomorphic to $\mathbb{Z}_2$.

---
[3] Using polynomial time Landau-Miller Test[18] it can be verified whether given polynomial has solvable Galois group or not.

Substituting $x = \frac{3}{10}(-9 - \sqrt{141})$ in the triangular form of the GB and solving for univariate polynomials in $y$ and $z$ we get, $y = \frac{1}{22}(3 - \sqrt{141}); z = \frac{1}{11}(14 + \sqrt{141})$, a sample solution. Galois group of irreducible polynomials of polynomial system are known a prior (isomorphic to $\mathbb{Z}_2$ for each variable $x, y, z$ over $\mathbb{Q}$). Rest of the polynomial solutions can be obtained by computing Galois-orbits of the sample solution.

In this example it is simple to observe the criterion (1) and (2) for deciding what solution tuple will form the equilibria.[4] Accepting values between 0 and 1 we get $x = \frac{3}{10}(-9 + \sqrt{141}); y = \frac{1}{22}(3 + \sqrt{141}); z = \frac{1}{11}(14 - \sqrt{141})$. This solution constitutes unique Nash equilibrium of the RPIE game depicted in Table 1. Note that, Galois group for the game system for this game is solvable, and so, all the equilibria computed are in closed form.

## 5   Computational Complexity

A Gröbner basis can be computed in doubly exponential time in the size of $\mathcal{K}^+$. A Gröbner basis contains polynomials in triangular form, and we are interested in the equilibria points with irrational values. An advantage of the triangular form is that at every stage of the substitution, unwanted solutions can be filtered out. For further details of computational complexity of finding Gröbner basis see [20].

The group action Algorithm 3.3, in the worst case, requires action of each of the Galois group generator $g' \in G' \subseteq G$ to each element of the set of roots [1]. This gives worst case time $O(|G'| \cdot |X|)$. If a univariate polynomial has $n$ roots then $|G'|$ is linear in $n$ [7,21], while $|X|$ is polynomial in $n$.

The algorithm for computing Nash equilibria with Gröbner basis, given in [4], substitutes all the roots in triangular form and solves univariate polynomial for each substitution. If Galois group is known for the polynomials then our approach computes solutions with relatively simple and efficient group action. The feature in our algorithm exploits information available in a sample solution. Our approach performs better then algorithm based only on Gröbner basis. If each univariate in Gröbner bases of the game system has $d_i$ distinct roots ($i \in \{1, \ldots, \mathcal{K}^+\}$), then method for computing Nash equilibria with only Gröbner basis takes $\Pi_i d_i$ substitutions and factorization. On the other hand, in our approach, after computing a sample solution, no further substitution or factorization is required.

## 6   Conclusion

We defined a class of rational payoff irrational equilibria games and proposed a purely algebraic method for computing its all equilibria. The method that we presented throws more light on structure of equilibria and shows a way to effectively utilizes knowledge of a sample equilibrium. We discussed results that show why our algorithm can not be used for any bimatrix games, and how it can be utilized to compute equilibria in closed form. Working of our algorithm was demonstrated with an example of 3 players 2 strategy RPIE game. For the class of games that we consider, the algorithm we suggest is new. It would be interesting to see working of our algorithm on graphical forms and extensive form games, with necessary modifications.

---

[4] For larger system the Nash equilibrium verification algorithm comes handy.